
\input vanilla.sty
\pagewidth{15.5 cm}
\pageheight{23 cm}
\magnification=\magstephalf
\TagsOnRight
\baselineskip 16pt
\define \la {\lambda}
\define \part {\partial}
\define \1 {\widetilde {\Cal V}^r_{(0)}}
\define \tc#1{$<A^{#1}\Phi_1,\Psi_1>-<A^{#1}\Phi_2,\Psi_2>$}
\define \ta#1{$<A^{#1}\Phi_1,\Psi_2>+<A^{#1}\Phi_2,\Psi_1>$}
\define \tb#1{$<A^{#1}\Phi_1,\Psi_2>-<A^{#1 }\Phi_2,\Psi_1>$}
\title
Binary nonlinearization for the Dirac systems\footnote"*"{
This work was supported by the National Natural Science
Foundation of China}
\endtitle
\author
Wen-xiu Ma \\
{\it Institute of Mathematics, Fudan University,
Shanghai 200433, P. R. of China}
\endauthor

\heading Abstract \endheading

{\narrower
A Bargmann symmetry constraint is proposed for the Lax pairs and the
adjoint Lax pairs of the Dirac systems. It is shown that the spatial
part of the nonlinearized Lax pairs and adjoint Lax pairs is a finite
dimensional Liouville integrable Hamiltonian system and that under the
control of the spatial part, the time parts of the nonlinearized Lax pairs
and adjoint Lax pairs are interpreted as a hierarchy of commutative,
finite dimensional Liouville integrable Hamiltonian systems whose
Hamiltonian functions consist of a series of integrals of motion for
the spatial part. Moreover an involutive representation of solutions
of the Dirac systems exhibits their integrability by quadratures.
This kind of symmetry constraint procedure involving the spectral problem
and the adjoint spectral problem is referred to as a binary nonlinearization
technique like a binary Darboux transformation.
\par}

{\narrower
\par}

\heading  1. Introduction \endheading

Symmetry constraints become prominent in recent few years due to the
important roles they play in soliton theory. For $1+1$ dimensional
soliton hierarchies, a very successful symmetry constraint method is
the nonlinearization technique  proposed by Cao$^{[1,2]}$ and Cao and
Geng$^{[3,4]}$. A large class of finite dimensional Liouville integrable
Hamiltonian systems is thus generated which are connected with
soliton hierarchies$^{[1-6]}$ and the nonlinearization technique  is
systematically extended by Zeng and Li$^{[7,8]}$ for quite a few
soliton hierarchies. At the same time,
a symmetry constraint procedure for bi-Hamiltonian
soliton hierarchies
is proposed by Antonowicz and Wojciechowski$^{[9]}$
and bi-Hamiltonian
structures for the constrained systems can be worked out$^{[9,10]}$.
By observing that stationary
soliton flows may be interpreted as finite dimensional Hamiltonian
systems$^{[11]}$ on introducing the so-called Jacobi-Ostrogradsky
coordinates$^{[12]}$,
a natural generalization of nonlinearization technique  to higher order
symmetry constraints is made by Zeng$^{[13]}$ for the KdV and
Boussinesq hierarchies.

For $1+2$ dimensional soliton hierarchies, a sort of symmetry
constraints similar to ones in the nonlinearization technique  has also
been presented by several authors$^{[14]}$,
although relatively
little is known about the general construction at present. Some
interrelations
of $1+2$ dimensional soliton hierarchies to $1+1$ dimensional soliton
ones and various structures of constrained systems have vigorously
been exposed$^{[15]}$.
The theory progress along the above two lines
has been making
and has aroused
increasing interest
in recent times.

This paper is devoted to the symmetry constraints in
the first line. Based upon the idea of
nonlinearization technique, we
would like to propose a binary
nonlinearization technique for soliton hierarchies possessing Lax pairs.
We shall express,
in terms of the eigenfunctions
and the adjoint  eigenfunctions,
the variational derivative of the spectral parameter
with respect to the potential,
and thus the binary
nonlinearization technique  involves two sets of dependent variables,
which is different from the
nonlinearization technique.
By considering the constraint problem of the Lax pairs and the
adjoint Lax pairs, we shall exhibit the idea of
binary nonlinearization technique and provide new examples for finite
dimensional Liouville integrable Hamiltonian systems.
In Section 2, we shall give the concrete construction of the Dirac
systems and their Hamiltonian structures.
Furthermore we want to establish some properties which will be used
later. In Section 3, we shall exhibit systematically the binary
nonlinearization procedure for the Lax pairs and the adjoint Lax
pairs of the Dirac systems.

\heading  2.
The Dirac systems and their Hamiltonian structures
\endheading

We consider the Dirac spectral problem$^{[16]}$
$$\phi _x=U\phi =U(u,\la )\phi ,\ \phi =
\pmatrix \phi _1 \\ \phi _2 \endpmatrix ,\ u =
\pmatrix q \\ r \endpmatrix,\tag2.1a$$
where the spectral operator is as follows
$$U=U(u,\la )=r \sigma_1 +\la i\sigma_2  +q \sigma_3
= \pmatrix q & \la + r \\ -\la + r  &- q  \endpmatrix,\tag2.1b$$
and the $\sigma _j ,\,1\le j\le 3,$ are $2\times 2$ Pauli matrices.
To derive Dirac systems associated with (2.1), we first solve the
adjoint representation equation (see Ref. [17]) $V_x=[U,V]$ of
$\phi _x=U\phi$. Set
$$V =a \sigma_1 +b i\sigma_2  +c \sigma_3   =\pmatrix c&a+b\\a-b&-c
\endpmatrix .\tag2.2$$
Noting that
$$[U,V]=(
-2\la c+2 q  b )\sigma_1 +(-2 r  c+2 q  a )i\sigma_2  +(2\la a -2 r  b )
\sigma_3 ,$$
we see that the adjoint representation equation $V_x=[U,V]$ becomes
$$\left \{\aligned &a_x=
-2\la c+2 q b ,\\
& b_x=-2 r c+2 q a ,
\\ & c_x=
2\la a -2 r b,\endaligned \right.$$
which is equivalent to
$$\left \{ \aligned
&a_0=c_0=0,\ b_{0x}=0,\\&
a_{ix}=-2c_{i+1}+2 q b_i, \\
&b_{ix}=-2rc_{i}+2qa_i,
\qquad i\ge 0,
\\
&c_{ix}=2a_{i+1}-2rb_i,
\endaligned \right.
\tag2.3$$
on setting
$a=\sum_{i\ge0}a_i\la ^{-i},\,
b=\sum_{i\ge0}b_i\la ^{-i},\,
c=\sum_{i\ge0}c_i\la ^{-i}.$
We choose
 $b_0=-1,$ and assume that
$a_i|_{u=0}=
b_i|_{u=0}=
c_i|_{u=0}=0,\ i\ge 1$ (or equivalently select constants of integration to
be zero). At this point, the recursion relation
(2.3) uniquely determines a series of differential
functions with respect to $u$.
For instance, we have
$$\align & a_1=-r,\ c_1=-q,\ b_1=0;\\
& a_2=-\frac12 q_x,\ c_2=\frac12 r_x,\ b_2=-\frac12 q^2-\frac12 r^2;\\
& a_3=\frac14 r_{xx}-\frac12 q^2r-\frac12 r^3,\
 c_3=\frac14 q_{xx}-\frac12 q^3-\frac12 q r^2,\
b_3=\frac12 (qr_x-q_x r).\endalign$$
{}From $(V^2)_x=[U, V^2]$, we see that
$(\frac 12 \text{tr}V^2)_x=(c^2+a^2-b^2)_x=0$. Thus by $
(\frac 12 \text{tr}V^2)|_{u=0}=1$,
we have $c^2+a^2-b^2=1$. Further we obtain
$$b_n=\frac12
\sum _{i=1}^{n-1}(b_ib_{n-i}
- a_ia_{n-i}
-c_ic_{n-i}) ,\ n\ge 2.$$
By the mathematical induction,
it follows from (2.3) and the above equality
that $a_i,b_i,c_i$ are all
differential polynomial functions of $u$.
The compatibility conditions of Lax pairs
$$\phi _x=U\phi ,\ \phi _{t_n}=V^{(n)}\phi ,\ V^{(n)}=(\la ^nV)_+,\ n\ge
0,\tag2.4$$
where the symbol $+$ stands for the selection of the polynomial part of
$\la $, engender a hierarchy of the Dirac systems
$$u_{t_n}=
\pmatrix q\\r\endpmatrix_{t_n}=K_n=
\pmatrix -2a_{n+1}\\2c_{n+1}\endpmatrix ,\ n\ge 0.\tag2.5$$
The first nonlinear Dirac system in the hierarchy (2.5)
reads as
$$\left \{\aligned &q_{t_2}=-\frac 12 r_{xx}+q^2r+r^3,\\
& r_{t_2}=\frac 12 q_{xx}-q^3-qr^2.\endaligned \right.$$
This system is different from the coupled nonlinear Schr\"odinger
systems in AKNS hierarchy because it contains the cubic terms $q^3,
r^3$.

In what follows, we want to give the Hamiltonian structures of the Dirac
systems (2.5) by means of the trace identity
proposed in Ref. [18].
To this end, we need the following quantities which are easy to
prove:
$$
<V, \frac {\part U} {\part \la}>= -2b,\
<V, \frac {\part U} {\part q}>= 2c,\
<V, \frac {\part U} {\part r}>= 2a,
\tag2.6 $$
where $<A,B>=\text{tr}(AB)$.
Now applying the trace identity$^{[18]}$
$$\frac \delta {\delta u}
<V, \frac {\part U} {\part \la}>= \la^{-\gamma}
\frac \part {\part \la}\la^\gamma (
<V, \frac {\part U} {\part q}>,
<V, \frac {\part U} {\part r}>
)^T,\ \gamma =\text{const.},$$
we obtain at once
$$\frac \delta {\delta u}(-2b)=
\la^{-\gamma } \frac \part {\part \la}\la^\gamma (2c,2a)^T.$$
Equating the coefficients of $\la^{-n-1}$ on two sides of the
above equality, we have
$$\frac \delta {\delta u}(-b_{n+1})=
(\gamma -n)(c_n,a_n,)^T,\  n\ge0. $$
By  taking simply $n=1$,
we find that the constant $\gamma =0$.
In this way we obtain an important
equality
$$\frac {\delta b_{n+1} }{\delta u}=
n(c_n,a_n,)^T,\  n\ge0.\tag2.7 $$
In addition, for $n\ge 1$ we have
$$
\pmatrix - 2a_{n+1}\\2c_{n+1}\endpmatrix=
\pmatrix 0&-1\\1&0\endpmatrix
\pmatrix 2c_{n+1}\\2a_{n+1}\endpmatrix=
\pmatrix 0&-1\\1&0\endpmatrix
\pmatrix
-2q\part ^{-1}r&
-\frac12 \part +2q\part ^{-1}q
\\
\frac12 \part -2r\part ^{-1}r
&2r\part ^{-1}q
\endpmatrix
\pmatrix 2c_{n}\\2a_{n}\endpmatrix.
$$
Therefore the hierarchy (2.5) may be cast into the following
Hamiltonian form
$$u_{t_n}=
\pmatrix q\\r\endpmatrix_{t_n}=K_n=
\pmatrix - 2a_{n+1}\\2c_{n+1}\endpmatrix=
JG_n=JL^n
\pmatrix 2c_1\\2a_1\endpmatrix=J
\frac {\delta H_n}{\delta u},\ n\ge 0,\tag2.8$$
where the Hamiltonian operator $J$, the recursive operator $L$ and
the Hamiltonian functions $H_n$ are determined by
 $$
J=
\pmatrix 0&-1\\1&0\endpmatrix,\ L=
\pmatrix
-2q\part ^{-1}r&
-\frac12 \part +2q\part ^{-1}q
\\
\frac12 \part -2r\part ^{-1}r
&2r\part ^{-1}q
\endpmatrix,\
H_n=\frac {2 b_{n+2}}{n+1},\  n\ge0.
\tag2.9 $$
By a direct calculation similar to Ref. [19], we can get
$$ V^{(m)}_{t_n}-V^{(n)}_{t_m}
+[V^{(m)},V^{(n)}]
=(V^{(m)})'[K_n]-(V^{(n)})'[K_m]
+[V^{(m)},V^{(n)}]
=0,$$ which implies the
commutability of the flows of (2.8).
Therefore each system in the hierarchy (2.8) has an infinite number
of symmetries $\{K_m\}_{m=0}^\infty$.
Besides, we can directly verify
$$V_{t_n}=[V^{(n)},V],\ n\ge 0,\tag2.10$$
when $u_{t_n}=K_n$, i.e. $U_{t_n}-V_x^{(n)}+[U,V^{(n)}]=0,\ n\ge0$.
In fact, we easily find that
$V_{t_n}-[V^{(n)},V]$ satisfies the adjoint representation equation
of $\phi _x=U\phi$ and that $V_{t_n}-
[V^{(n)},V]$ vanishes at $u=0$.
Thus (2.10) holds for $n\ge 0$ because the adjoint representation
equation $V_x=[U,V]$ has uniqueness, namely if $
V_x=[U,V]$ and $V$ vanishes at $u=0$, then $V$ itself vanishes.

\heading 3. Binary nonlinearization \endheading

Let us begin with the binary nonlinearization of the Lax pairs and
adjoint Lax pairs of the Dirac systems. Associated with Lax pairs
(2.4), the Dirac systems have the adjoint Lax pairs
\vskip2mm
\line {\hbox to 0pt {\hss}
\hss $\displaystyle \left \{\matrix \format \l\\
\psi _{x}=-U^T \psi =-U^T(u,\la )\psi,
\\ \vspace {2mm}
\psi _{t_n}=-(V^{(n)})^T \psi =-(V^{(n)})^T(u,\la )\psi,
\endmatrix
\right. $\hss
\hbox to 0pt {\hss $\displaystyle \matrix (3.1a)\\
\vspace {2mm} (3.1b)\endmatrix $}}
\vskip2mm
\noindent where $T$ means the transpose of matrix and $\psi
=(\psi_1,\psi_2)^T$.
It follows from $\phi_x=U\phi,\ \psi_x=-U^T\psi$, that
$$
\frac {\delta \la }{\delta u}
=\frac 1 E (\phi_1\psi_1-\phi_2
\psi_2,\phi_1\psi_2+\phi_2\psi_1)^T, \  E=
\int ^\infty_{-\infty}(\phi _{1}\psi_{2}-\phi_2\psi_1) \,dx.\tag3.2$$
When the
zero boundary conditions:
$\lim_{|x|\to +\infty}\phi=
\lim_{|x|\to +\infty}\psi=0,$ hold, we have
$$L \frac {\delta \la }{\delta u}=\la
\frac {\delta \la }{\delta u},\ \ \text{or}\ \ L^*J\frac {\delta \la }
{\delta u}=\la J\frac {\delta \la }{\delta u},\tag3.3$$
where $L$ is defined
as in (2.9), and $L^*$ is the adjoint operator of $L$, which is a
hereditary symmetry.
For a general spectral problem, this kind of
property (3.3) has been discussed in Ref. [20].

Now introducing $N$ distinct eigenvalues $\la _1,\la _2,\cdots,\la _N,$
we obtain the following two spatial and time systems
\vskip2mm
\line {\hbox to 0pt {\hss}
\hss $\displaystyle \left \{\matrix \format \l \\
\pmatrix \phi _{1j}\\ \phi _{2j}\endpmatrix
_x=U(u,\la _j)
\pmatrix \phi _{1j}\\ \phi _{2j}\endpmatrix ,\ j=1,2,\cdots,N,\\
\vspace {3mm}
\pmatrix \psi_{1j}\\ \psi_{2j}\endpmatrix
_x=-U^T(u,\la _j)
\pmatrix \psi_{1j}\\ \psi_{2j}\endpmatrix ,\ j=1,2,\cdots,N;
\endmatrix
\right. $\hss
\hbox to 0pt {\hss $\displaystyle \matrix (3.4a)\\
\vspace {8mm} (3.4b)\endmatrix $}}
\vskip2mm
\line {\hbox to 0pt {\hss}
\hss $\displaystyle \left \{\matrix \format \l \\
\pmatrix \phi _{1j}\\ \phi _{2j}\endpmatrix
_{t_n}=V^{(n)}(u,\la _j)
\pmatrix \phi _{1j}\\ \phi _{2j}\endpmatrix ,\ j=1,2,\cdots,N,\\
\vspace {3mm}
\pmatrix \psi_{1j}\\ \psi_{2j}\endpmatrix
_{t_n}=-(V^{(n)})^T(u,\la _j)
\pmatrix \psi_{1j}\\ \psi_{2j}\endpmatrix ,\ j=1,2,\cdots,N.
\endmatrix
\right. $\hss
\hbox to 0pt {\hss $\displaystyle \matrix (3.5a)\\
\vspace {8mm} (3.5b)\endmatrix $}}
\vskip2mm
\noindent Because
$U_{t_n}-V^{(n)}+[U,V^{(n)}]=0$ if and only if
$(-U^T)_{t_n}-(-(V^{(n)})^T)_x+[-U^T,-(V^{(n)})^T]=0$,
the compatibility condition of (3.4) and (3.5) is still
the $n$th Dirac
systems $u_{t_n}=K_n$.
Let us take the Bargmann constraint (requiring the $G$-vector field to be
a linear function but not a differential function with respect to $u$)
for the Dirac systems
$$JG_0=J\sum_{j=1}^N2E_j
\frac {\delta \la _j }{\delta u},\
E_j=\int ^\infty_{-\infty}(\phi _{1j}\psi_{2j}-\phi_{2j}\psi_{1j})
\,dx.\tag3.6$$
This kind of constraints is, in fact, symmetry constraints because the
$J\dfrac {\delta \la _j}{\delta u}$ are common symmetries of the Dirac
systems.
The constraint (3.6) allows us to impose that
$$\pmatrix c_1\\a_1\endpmatrix =\bigl (
<\Phi_1,\Psi_1>-<\Phi_2,\Psi_2>,
<\Phi_1,\Psi_2>+<\Phi_2,\Psi_1>\bigr)^T,$$
which implies that
\vskip2mm
\line {\hbox to 0pt {\hss}
\hss $\displaystyle \left \{\matrix \format \l \\
q=- <\Phi_1,\Psi_1>+<\Phi_2,\Psi_2>,\\ \vspace {2mm}
r=- <\Phi_1,\Psi_2>-<\Phi_2,\Psi_1>.
\endmatrix\right. $\hss
\hbox to 0pt {\hss $\displaystyle \matrix (3.7a)\\
\vspace {2mm} (3.7b)\endmatrix $}}
\vskip2mm
\noindent Here $\Phi_i=(\phi_{i1},\cdots,\phi_{iN})^T,
\ \Psi_i=(\psi_{i1},\cdots,\psi_{iN})^T,$ $i=1,2$, $<y,z>=\sum
_{j=1}^Ny_jz_j,\, y,z\in R^N$. Note that (3.7) has two particular
properties: including both the eigenfunctions $\Phi_1,\Phi_2$
and the adjoint
eigenfunctions $\Psi_1,\Psi_2$, and nonlinearity with respect
to $\Phi_i,\Psi_i$. Therefore we refer to (3.7) as a binary nonlinear
constraint. In general, nonlinear constraints only involve
eigenfunctions
of spectral problems associated with integrable systems (see Refs.
[1-8]). We denote by $\widetilde  A$ the expression of $A$ under the
constraint (3.7).
The property (3.3) ensures that
$$\pmatrix \widetilde  c_n \\
\vspace {2mm}
\widetilde a _n \endpmatrix =\widetilde  L ^{n-1}
\pmatrix \widetilde  c_1 \\
\vspace {2mm}
\widetilde  a _1 \endpmatrix
=\pmatrix \text{\tc {n-1}}  \\
\vspace {2mm}
\text{\ta {n-1}}\endpmatrix,\ n\ge 1,\tag3.8a$$
and that from (2.3),
$$\widetilde  b_n=\part ^{-1}(2\widetilde  q \,\widetilde  a_n-2\widetilde
r\,\widetilde  c_n)=\text{\tb {n-1}},\ n\ge 1.\tag3.8b$$
Here $\widetilde  b_1\ne 0$. But $\widetilde  V_x=[\widetilde
U,\widetilde V]$ still holds. By
substituting (3.7) into the Lax pairs and the adjoint Lax pairs:
(3.4) and (3.5), we acquire the
nonlinearized Lax pairs and adjoint Lax pairs
\vskip2mm
\line {\hbox to 0pt {\hss}
\hss $\displaystyle \left \{\matrix \format \l \\
\pmatrix \phi _{1j}\\ \phi _{2j}\endpmatrix
_x=U(
\widetilde  u,\la _j)
\pmatrix \phi _{1j}\\ \phi _{2j}\endpmatrix ,\ j=1,2,\cdots,N,\\
\vspace {3mm}
\pmatrix \psi_{1j}\\ \psi_{2j}\endpmatrix
_x=-U^T(
\widetilde  u,\la _j)
\pmatrix \psi_{1j}\\ \psi_{2j}\endpmatrix ,\ j=1,2,\cdots,N;
\endmatrix
\right. $\hss
\hbox to 0pt {\hss $\displaystyle \matrix (3.9a)\\
\vspace {8mm} (3.9b)\endmatrix $}}
\vskip2mm
\line {\hbox to 0pt {\hss}
\hss $\displaystyle \left \{\matrix \format \l \\
\pmatrix \phi _{1j}\\ \phi _{2j}\endpmatrix
_{t_n}=V^{(n)}(
\widetilde  u,\la _j)
\pmatrix \phi _{1j}\\ \phi _{2j}\endpmatrix ,\ j=1,2,\cdots,N,\\
\vspace {3mm}
\pmatrix \psi_{1j}\\ \psi_{2j}\endpmatrix
_{t_n}=-(V^{(n)})^T(
\widetilde  u,\la _j)
\pmatrix \psi_{1j}\\ \psi_{2j}\endpmatrix ,\ j=1,2,\cdots,N.
\endmatrix
\right. $\hss
\hbox to 0pt {\hss $\displaystyle \matrix (3.10a)\\
\vspace {8mm} (3.10b)\endmatrix $}}
\vskip2mm
\noindent The spatial part of the nonlinearized
Lax pairs and adjoint Lax pairs, i.e.
the system (3.9)
is a finite dimensional system
but
the time parts of the
nonlinearized Lax pairs and adjoint Lax pairs, i.e.
the systems
(3.10) for $n\ge 0$ are all systems of evolution equations
in $1+1$ dimensions.
In the following, we shall verify that the system (3.9) is a
Liouville integrable (see Ref. [21]) Hamiltonian system and that under
the control of (3.9), the systems (3.10) are also Liouville
integrable Hamiltonian systems.

The system (3.9) may be cast into the following Hamiltonian form
$$\Phi_{ix}=\frac {\part H}{\part \Psi_{i}},\ \Psi_{ix}=-\frac {\part
H}{\part \Phi_i},\ i=1,2.\tag3.11a$$
where the Hamiltonian function
$$\align H=&-\frac12 \bigl ( <\Phi
_1,\Psi_1>-<\Phi_2,\Psi_2>\bigr)^2\\&
-\frac12 \bigl ( <\Phi _1,\Psi_2>+<\Phi_2,\Psi_1>\bigr)^2-
< A\Phi _1,\Psi_2>+<A\Phi_2,\Psi_1>.\tag3.11b\endalign$$
Let us now construct integrals of motion for (3.11).
An obvious equality $(\widetilde  V^2)_x=[\widetilde U,\widetilde  V^2]$
leads to
$$F_x=(\frac12 \text{tr}\widetilde  V^2)_x=\frac d {dx}(
\widetilde  c^2+
\widetilde  a^2-
\widetilde  b^2)=0.$$
Thus $F$ is a generating function of
integrals of motion for (3.11).
Since $F=\sum_{n\ge 0}F_n\la ^{-n}$, we obtain the following
expressions
$$F_n=\sum_{i=0}^n (
\widetilde  c_i \widetilde  c_{n-i}
+\widetilde  a_i \widetilde  a_{n-i}-
\widetilde  b_i \widetilde  b_{n-i}).$$
Further by (3.8), we have
$$\align F_0=&-1,\
F_1=2\bigl(<\Phi_1,\Psi_2>-<\Phi_2,\Psi_1>\bigr),\tag3.12a\\
F_n=&\sum_{i=1}^{n-1}(
\widetilde  c_i \widetilde  c_{n-i}
+\widetilde  a_i \widetilde  a_{n-i}-
\widetilde  b_i \widetilde  b_{n-i})-2\widetilde  b_0\widetilde  b_n\\
=&
\sum_{i=1}^{n-1}\Big [\big(\text{\tc {{i-1}}} \bigr)
\bigl( \text{\tc {{n-i-1}}} \bigr)\\
\vspace {1mm}
&+
\bigl( \text{\ta {{i-1}}} \bigr)
\bigl( \text{\ta {{n-i-1}}} \bigr)\\
\vspace {1mm}
&-
\bigl( \text{\tb{{i-1}} }\bigr)
\bigl( \text{\tb {{n-i-1}}} \bigr)\Bigr]\\
\vspace {1mm}
&+
2\bigl(\text{\tb {{n-1}} }\bigr),\ n\ge 2.\tag3.12b\endalign$$
Here $F_n,\, n\ge0,$ are all polynomials of $4N$ dependent variables
$\phi_{ij},\, \psi_{ij}$.
Note that we have (2.10). With the same deduction, we find that
$F=\frac12 \text{tr}\widetilde  V^2$ is also a generating function of
integrals of motion for (3.10). Moreover, we have
$$\align &\bigl ( \frac {\part F}{\part \Phi_1},
\frac {\part F}{\part \Phi_2}\bigr)^T =\Bigl (
\text{tr} \big
(\widetilde  V\frac \part {\part \Phi_1}\widetilde  V\bigr),
\text{tr} \big
(\widetilde  V\frac \part {\part \Phi_2}\widetilde  V\bigr)\Bigr)^T,\tag3.13\\
 &\text{tr} \big
(\widetilde  V\frac \part {\part \Phi_1}\widetilde  V\bigr)=
\text{tr}\sum_{i=0}^\infty
\pmatrix \widetilde  c_i &\widetilde  a_i +\widetilde  b_i\\
\widetilde  a_i-\widetilde  b_i&-\widetilde  c_i\endpmatrix\la ^{-i}
\sum_{j=0}^\infty\frac \part{\part \Phi_1}
\pmatrix \widetilde  c_j &\widetilde  a_j +\widetilde  b_j\\
\widetilde  a_j-\widetilde  b_j&-\widetilde  c_j\endpmatrix\la ^{-j}
\\&=\text{tr}\sum _{\Sb i\ge 0\\ \vspace {1mm}j\ge1\endSb}
\pmatrix \widetilde  c_i &\widetilde  a_i +\widetilde  b_i\\
\widetilde  a_i-\widetilde  b_i&-\widetilde  c_i\endpmatrix
\pmatrix A^{j-1}\Psi_1&2A^{j-1}\Psi_2\\0&-A^{j-1}\Psi_1\endpmatrix
\la ^{-(i+j)}\\&=
2\sum _{\Sb i\ge 0\\ \vspace {1mm}j\ge1\endSb}
\bigl [ \widetilde  c_i A^{j-1}\Psi_1+(\widetilde  a_i-\widetilde
b_i)A^{j-1}\Psi_2\bigr]
\la ^{-(i+j)}.\tag3.14a\endalign $$
Similarly we can obtain
$$
\text{tr} \big
(\widetilde  V\frac \part {\part \Phi_2}\widetilde  V\bigr)=
2\sum _{\Sb i\ge 0\\ \vspace {1mm}j\ge1\endSb}
\bigl [ (\widetilde  a_i+\widetilde  b_i) A^{j-1}\Psi_1
-\widetilde  c_i A^{j-1}\Psi_2\bigr]
\la ^{-(i+j)}.\tag3.14b$$
The equalities (3.13) and (3.14) lead to
$$\frac12 \pmatrix \dfrac {\part F_{n+1}}{\part \Phi_1}\\
\vspace {1mm}
\dfrac {\part F_{n+1}}{\part \Phi_2}\endpmatrix =-(-\widetilde
V^{(n)})^T\pmatrix \Psi_1\\ \Psi_2\endpmatrix,\ n\ge0.$$
The similar deduction can give rise to
$$\frac12 \pmatrix \dfrac {\part F_{n+1}}{\part \Psi_1}\\
\vspace {1mm}
\dfrac {\part F_{n+1}}{\part \Psi_2}\endpmatrix =\widetilde
V^{(n)}\pmatrix \Phi_1\\ \Phi_2\endpmatrix,\ n\ge0.$$
At this stage, the systems (3.10) may be readily rewritten as
$$\Psi_{it_n}=-\frac {\part (\frac12 F_{n+1})}{\part \Phi_i},\
\Phi_{it_n}=\frac {\part (\frac12 F_{n+1})}{\part \Psi_i},\
i=1,2. \tag3.15 $$
Associated with the symplectic structure on $R^{4N}$
$$\omega ^2=d\Psi_1\wedge d\Phi_1+d\Psi_2\wedge d\Phi_2,$$
we can generate the Poisson bracket
$$\{P,Q\}=\omega ^2 (IdQ,IdP),\tag3.16$$
where $IdR$ denotes a Hamiltonian vector field of a smooth function
$R$ on $R^{4N}$, defined by $IdR\text{\_\kern-.0em l}\,\omega^2=-dR$,
with $\text{\_
\kern-.25em l}$ being the left interior product.
Since $F$ is a generating function
of integrals of motion for (3.10), we obtain
$$\{F_{m+1},F_{n+1}\}=2\frac \part {\part t_n}F_{m+1}=0,\ m,n\ge
0.\tag3.17$$
This elucidates that $\{F_n\}_{m=0}^\infty$ constitutes a Poisson algebra
with regard to (3.16).
Evidently we have
$$
\left. \frac {\part F_n}{\part \Phi _{1}}\right|_{\Phi _1=\Phi
_2=0}=2A^{n-1}\Psi _2,\
\left. \frac {\part F_n}{\part \Phi _{2}}\right|_{\Phi _1=\Phi
_2=0}=-2A^{n-1}\Psi _1,\ n\ge 1.\tag3.18$$
Therefore
there must exist one region $\Omega \subseteq R^{4N}$ on which the
$2N$ 1-forms $dF_1,\cdots, dF_{2N}$ are every linearly independent
since the Vandermonde determinant $V(\la _1,\la_ 2,\cdots,\la _N)$
is nonzero.
In this way we have shown that the spatial part of
the nonlinearized Lax
pairs and adjoint Lax pairs and the time parts of
the nonlinearized Lax
pairs and adjoint Lax pairs under the control of the spatial part are all
finite dimensional integrable Hamiltonian systems in the Liouville
sense (see Ref. [21]).

When
$\lim_{|x|\to +\infty}\Psi_i=
\lim_{|x|\to +\infty}\Phi_i=0,$ $i=1,2$ (at this stage, $\widetilde  b_1=
\frac12 F_1=0$ since $F_{1x}=0$)
and $\Phi_i(x,t_n),\,\Psi_i(x,t_n)$ solve simultaneously
(3.11) and (3.15),
then
$$u=\bigl ( -<\Phi_1,\Psi_1>+<\Phi_2,\Psi_2>,-<\Phi_1,\Psi_2>-<\Phi_2,
\Psi_1> \bigr)^T$$
is a solution to the $n$th Dirac system
$u_{t_n}=K_n$. This allows us to conclude that the $n$th Dirac system
$u_{t_n}=K_n$ has the following involutive representation of
solutions
\vskip2mm
\line {\hbox to 0pt {\hss}
\hss $\displaystyle \left \{\matrix
q(x,t_n)=-<
g^x_Hg^{t_n}_{F_{n+1}}\Phi_1(0,0),
g^x_Hg^{t_n}_{F_{n+1}}\Psi_1(0,0)>\\
\vspace {2mm}
\qquad\qquad
+<g^x_Hg^{t_n}_{F_{n+1}}\Phi_2(0,0),
g^x_Hg^{t_n}_{F_{n+1}}\Psi_2(0,0)>,\\
\vspace {2mm}
r(x,t_n)=-
<g^x_Hg^{t_n}_{F_{n+1}}\Phi_1(0,0),
g^x_Hg^{t_n}_{F_{n+1}}\Psi_2(0,0)>\\
\vspace {2mm}
\qquad\qquad
-<g^x_Hg^{t_n}_{F_{n+1}}\Phi_2(0,0),
g^x_Hg^{t_n}_{F_{n+1}}\Psi_1(0,0)>,
\endmatrix
\right. $\hss
\hbox to 0pt {\hss $\displaystyle \matrix  \  \\ \vspace {4mm}(3.19a)\\
\vspace {2mm} \ \\ \vspace {3mm}(3.19b)\endmatrix $}}
\vskip2mm
\noindent with $g^x_H,\,g^{t_n}_{F_{n+1}}$ being the Hamiltonian phase
flows (see Ref.[21]) associated with the Hamiltonian functions
$H,\,F_{n+1}$, but $\Phi_i(0,0),\,\Psi_i(0,0),\, i=1,2$, being any
fixed constant vector of $R^N$. This kind of involutive
representation of solutions to the Dirac systems also show, to some extent,
the characteristic of integrability of the Dirac systems in view of the
Liouville integrability of the flows $g_H^x,\,g^{t_n}_{F_{n+1}}$.

\heading 4. Conclusions and remarks \endheading

We introduced a kind of symmetry constraint on the Dirac integrable
systems. Moreover we exhibited an explicit Poisson algebra on the
symplectic manifold $(R^{4N},\omega ^2)$ and further, an involutive
representation of solutions to the Dirac systems.
The proposed symmetry constraint includes the eigenfunctions and the adjoint
eigenfunctions and is nonlinear with respect to them. The
corresponding reductions of the Lax pairs and adjoint Lax pairs and the
manipulation with the integrability of the constrained systems
constitute a binary nonlinearization
problem.

We remark that when the zero boundary conditions on
the eigenfunctions and the adjoint
eigenfunctions are not imposed, the spatial part of
the nonlinearized Lax pairs and adjoint Lax pairs
is invariant and thus, still a Hamiltonian system with the original
Hamiltonian function. However, the time parts of
the nonlinearized Lax pairs and adjoint Lax pairs
will vary with the boundary conditions.
When the zero boundary conditions are not satisfied, they become very
complicated Hamiltonian systems. Nevertheless, their Hamiltonian functions
are some polynomials of $F_n,\,n\ge1$, which may be determined
by an obtained equality $c^2+a^2-b^2=1$.
Furthermore we can consider higher order constraints
$$JG_n=J\sum _{j=1}^N2E_j\frac {\delta \la _j}{\delta u},\ (n\ge1).$$
This kind of symmetry constraints is somewhat different from the
Bargmann constraint because the $G_n$ involve the differential of $u$ with
respect to $x$. To discuss them, we need to introduce new dependent variables,
i.e. the so-called Jacobi-Ostrogradsky coordinates.
These problems are left to a further investigation.

There exist also certain relations between
the nonlinearization technique and the binary
nonlinearization technique. For example, if
we take the possible reduction
$\Psi_1=-\Phi_2,\,\Psi_2=\Phi_1$, the involutive system (3.12) simplifies to
$$
\align  F_1=&2\bigl(<\Phi_1,\Phi_1>+<\Phi_2,\Phi_2>\bigr),\\
\vspace {1mm}
F_n=&\sum_{i=1}^{n-1}
\Big [
\bigl( <A^{i-1}\Phi_1,\Phi_1>-<A^{i-1}\Phi_2,\Phi_2> \bigr)
\bigl( <A^{n-i-1}\Phi_1,\Phi_1>-<A^{n-i-1}\Phi_2,\Phi_2> \bigr)
\Bigr.
\\
\vspace {1mm}
&-
\bigl( <A^{i-1}\Phi_1,\Phi_1>+<A^{i-1}\Phi_2,\Phi_2>\bigr)
\bigl( <A^{n-i-1}\Phi_1,\Phi_1>+<A^{n-i-1}\Phi_2,\Phi_2>\bigr)
\\
\vspace {1mm}
&
\Bigl.+4<A^{i-1}\Phi_1,\Phi_2><A^{n-i-1}\Phi_1,\Phi_2>
\Bigr]
+
2\bigl(<A^{n-1}\Phi_1,\Phi_1>+<A^{n-1}\Phi_2,\Phi_2>
\bigr),\ n\ge 2.\endalign$$
This is an
involutive system on the symplectic manifold $(R^{2N}, \omega
^2=d\Phi_1\wedge d\Phi_2)$, and it
may be generated by the nonlinearization technique.

We should note that the idea of binary nonlinearization is quite
broad; it can be applied to other integrable systems (see, for example,
Ref. [22]).
Therefore a large class of finite dimensional Liouville integrable
Hamiltonian systems may be raised by means of our binary
nonlinearization technique and the involutive representation of
solutions can exhibit the integrability by quadratures for the
considered systems.

\vskip 0.5cm
\noindent {\bf
Acknowledgements:}
The author expresses his gratitude to Prof.
Gu C H and Prof. Hu H S for continuous support during the preparation of
the paper. The author is also grateful to Profs.  Cao C W, Zeng Y B
and Geng X G for stimulating
discussions.

\heading References \endheading
\item {[1]}
Cao C W, 1987, Henan Sci. 5, 1;
1988, Chin. Quart. J. Math. 3, 90.
\item {[2]} Cao C W, 1990,
Acta Math. Sinica New Series  6, 35;
1990, Sci. China A 33, 528.
\item {[3]}
 Cao C W and Geng X G, 1990,
In: Nonlinear Physics, eds. Gu C H,  Li Y S and Tu G Z
(Berlin: Springer-Verlag) p.68.
\item {[4]}
Cao C W and Geng X G,
1990, J. Phys. A 23, 4117; 1991, J. Math. Phys. 32, 2323.
\item {[5]} Zhuang D W and Lin Y Q, 1990,
In: Nonlinear Physics, eds. Gu C H,  Li Y S and Tu G Z
(Berlin: Springer-Verlag)
p.92;
 Xu T X and Mu W H, 1990, Phys. Lett. A 147, 125;
 Gu Z Q, 1991, Chin. Sci. Bull.  36, 1683;
Geng X G, 1992, Physica A 180, 241;
 Qiao Z J, 1993, Phys. Lett. A  172, 224.
\item {[6]}
Ma W X, 1990, In: Nonlinear Physics eds. Gu C H,  Li Y S and Tu G Z
(Berlin: Springer-Verlag) p.79;
1990, Chin. Sci. Bulletin 35, 1853;
1992, Appl. Math. Mech. 13, 369;
1993, Acta Math. Appl. Sinica  9, 92.
\item {[7]}
Zeng Y B and Li Y S, 1989, J. Math. Phys.  30, 1679;
1990, Acta Math. Sinica New Series 6, 257.
\item {[8]}
Zeng Y B and  Li Y S,
1990, J. Phys. A: Math. Gen. 23, L89;
1990, J. Math. Phys. 31, 2835.
\item {[9]}
Antonowicz M and Wojciechowski S,
1990, Phys. Lett. A  147, 455;
1991, J. Phys A: Math. Gen. 24, 5043;
1992, J. Math. Phys. 33, 2115.
\item {[10]} Blaszak M,
1993, Phys. Lett. A 174, 85;
1993, J. Phys. A: Math. Gen. 26, 5985.
\item {[11]} Bogoyavlenskii O I and Novikov S P,
1976, Funct. Anal. Appl.
10, 8; Flaschka H, Newell A C and Ratiu T, 1983, Physica D
9, 324;
Antonowicz M, Fordy A P and Wojciechowski S,
1987, Phys. Lett. A 124, 143;
 Tu G Z, 1991, Advances in Math. (China) 20, 60.
\item {[12]} Dele\'on M and Rodrigues P R, 1985, Generalized
Classical Mechanics and Field Theory
(Amsterdam: North-Holland).
\item {[13]}
Zeng Y B,
1991, Phys. Lett. A  160, 541; 1992, Chin. Sci. Bull. 37, 769.
\item {[14]}
Konopelchenko B, Sidorenko J and Strampp W, 1991, Phys.
Lett. A 157, 17;
Cheng Y and Li Y S, 1991, Phys. Lett. A {\bf 157},
22;
Konopelchenko B and Strampp W, 1991,
Inverse Problems 7, L17;
1992, J. Math. Phys. 33, 3676;
Sidorenko J and Strampp W, 1991,
Inverse Problems
 7, L37;
Cheng Y, 1992, J. Math. Phys. 33, 3774.
\item {[15]}
Xu B and Li Y S, 1992,
J. Phys. A: Math. Gen.
 25, 2957;
Xu B, 1992, Inverse Problems  8, L13;
 Oevel W
and Strampp W, 1993,
Commun. Math. Phys. 157, 51; Liu Q P, 1994,
Phys. Lett. A 187, 373.
\item {[16]} Frolov I S, 1972, Sov. Math. Dokl. 13, 1468;
Hinton D B,  Jordan A K, Klaus M and Shaw J K,
1991, J. Math. Phys. 32, 3015;
Xu B Z and Jiang X F, 1993, Hamiltonian structures for two sorts of evolution
equations, preprint.
\item {[17]}
Fordy A P and Gibbons J, 1980, J. Math. Phys. 21, 2508;
1981, ibid, 22, 1170.
\item {[18]}
Tu G Z, 1986,
Scientia Sinica  24, 138; 1989,
J. Math. Phys. 33, 330;
Tu G Z and Meng D Z, 1989, Acta Math. Appl. Sinica 5, 89.
\item {[19]}
Ma W X, 1992, J. Math. Phys.
33, 2464; 1992, J. Phys. A: Math. Gen. 25, 5329; 1993,
Differential Geometric Methods in Theoretical Physics
(Singapore: World Scientific) p.535.
\item {[20]} Fokas A S and Anderson R L, 1980, J. Math. Phys.
23, 1066; Tu G Z, 1990, Northeastern Math. J. 6, 26.
\item {[21]} Arnold V I, 1978, Mathematical Methods of Classical
Mechanics (Berlin: Springer-Verlag);
Abraham R and Marsden J, 1978, Foundations of Mechanics 2nd ed.
(Massachusetts: Addison-Wesley).
\item {[22]} Ma W X, 1995, J. Phys. Soc. Jpn. 64, 1085;
Ma W X and Strampp W,
1994, Phys. Lett. A 185, 277.
\bye